\documentclass{IEEEtran4PSCC}
\ifCLASSINFOpdf
   \usepackage[pdftex]{graphicx}
\else
   \usepackage[dvips]{graphicx}
\fi
\usepackage{cite}
\usepackage{mathtools}
\usepackage{adjustbox}
\usepackage{amssymb,amsfonts,amsmath}
\usepackage{algorithm}
\usepackage{algorithmic}
\usepackage{graphicx,multirow}
\usepackage{textcomp}
\usepackage{color,soul}
\usepackage{enumerate}
\usepackage{pgfplots}
\usepackage{physics}
\usepackage{url}
\usepackage{booktabs,colortbl}
\usepackage{array}
\usepackage{optidef}
\usepackage{bm}
\usepackage{dcolumn}
\usepackage{longtable}
\usepackage{makecell}
\usepackage{adjustbox}
\usepackage[caption=false, font=footnotesize]{subfig}
\makeatletter
\let\old@ps@headings\ps@headings
\let\old@ps@IEEEtitlepagestyle\ps@IEEEtitlepagestyle
\def\psccfooter#1{%
    \def\ps@headings{%
        \old@ps@headings%
        \def\@oddfoot{\strut\hfill#1\hfill\strut}%
        \def\@evenfoot{\strut\hfill#1\hfill\strut}%
    }%
    \def\ps@IEEEtitlepagestyle{%
        \old@ps@IEEEtitlepagestyle%
        \def\@oddfoot{\strut\hfill#1\hfill\strut}%
        \def\@evenfoot{\strut\hfill#1\hfill\strut}%
    }%
    \ps@headings%
}
\makeatother

\begin{document}
%
\title{Predicting Electricity Infrastructure Induced Wildfire Risk in California}

\author{
\IEEEauthorblockN{Mengqi Yao\\ Duncan S. Callaway}
\IEEEauthorblockA{Energy and Resources Group\\
University of California, Berkeley\\
Berkeley, USA\\
\{mqyao, dcal\}@berkeley.edu}
}

 \author{\IEEEauthorblockN{Mengqi Yao\IEEEauthorrefmark{1},
 Meghana Bharadwaj\IEEEauthorrefmark{2},
 Zheng Zhang\IEEEauthorrefmark{3}, 
 Baihong Jin\IEEEauthorrefmark{2} and
 Duncan S. Callaway\IEEEauthorrefmark{1}}
\IEEEauthorblockA{\IEEEauthorrefmark{1}Energy and Resources Group, University of California, Berkeley, CA, USA}
\IEEEauthorblockA{\IEEEauthorrefmark{2}Department of Electrical Engineering and Computer Sciences, University of California, Berkeley, CA, USA}
\IEEEauthorblockA{\IEEEauthorrefmark{3}College of Letters and Science, University of California, Berkeley, CA, USA}
}

\maketitle

\begin{abstract}
This paper examines the use of risk models to predict the timing and location of wildfires caused by electricity infrastructure.  Our data include historical ignition and wire-down points triggered by grid infrastructure collected between 2015 to 2019 in Pacific~Gas~\&~Electricity territory along with various weather, vegetation, and very high resolution data on grid infrastructure including location, age, and materials. With these data we explore a range of machine learning methods and strategies to manage training data imbalance. The best area under the receiver operating characteristic we obtain is 0.776 for distribution feeder ignitions and 0.824 for transmission line wire-down events, both using the histogram-based gradient boosting tree algorithm (HGB) with under-sampling. We then use these models to identify which information provides the most predictive value. After line length, we find that weather and vegetation features dominate the list of top important features for ignition or wire-down risk.  Distribution ignition models show more dependence on slow-varying vegetation variables such as burn index, energy release content, and tree height, whereas transmission wire-down models rely more on primary weather variables such as wind speed and precipitation.  These results point to the importance of improved vegetation modeling for feeder ignition risk models, and improved weather forecasting for transmission wire-down models.  We observe that infrastructure features make small but meaningful improvements to risk model predictive power. 
\end{abstract}

\vspace{2pc}
\noindent{\it Keywords}: Wildfire, risk prediction, machine learning, imbalance classification, feature importance


\section{Introduction}

Although wildfire ignitions triggered by the power grid in California only count 8\% of events in California~\cite{syphard2015location}, these ignitions often result in large fires, causing extensive damage because they coincide with hot and windy conditions. For example, California's 2018 Camp Fire, in which the failure of a transmission line ignited a fire that killed 84 people, resulted in \$9.3 billion in housing damage, and eventually led Pacific~Gas~\&~Electricity (PG\&E) to file for bankruptcy~\cite{wildfire_2019}.  As reported in~\cite{WMP_2021}, PG\&E calculates an average cost of ignition to be \$5.2 million. 

Most existing wildfire risk modeling research focuses on environment-caused wildfires \cite{westerling2018wildfire,sakr2010artificial} or human-caused wildfires~\cite{rodrigues2014insight}, and few focus on those cause by power grids. Previous studies using statistical methods only investigated correlation between a small number of environmental features (i.e., wind speed, fuel moisture) and the ignitions triggered by power lines~\cite{mitchell2013power}. Weather and vegetation data were commonly used in past work~\cite{lall2016application,storer2016pso}, but grid infrastructure data has been minimally investigated. In~\cite{malik2021data}, only the location of transmission lines was included, and the study considered a small area near Monticello and Winters, California. Previous work on wildfire risk estimation in power system overlayed power lines on a general fire risk map to identify high-risk power lines~\cite{rhodes2020balancing,hong2022data}, without considering the specific wildfire risk of electric components.  While some utilities have built their own wildfire risk models, there are no such models in the public domain.

The objective of this paper is to understand what data are most valuable for ignition prediction and to provide recommendations for what types of data should be gathered, improved upon, and incorporated into risk models.  We use PG\&E as a case study due to its history with wildfire and widely available data. We employ a set of machine learning models that use a range of weather, vegetation, and infrastructure characteristics to predict the risk of grid-induced ignitions and wire-down events.  { Here, we only predict the probability of an ignition.  While the consequences of ignition -- for example burned acres, structures destroyed, and outage duration -- are of clear societal importance, they depend strongly on post-ignition factors such as fire weather and human decision-making regarding fire suppression.  Best practice in this wildfire spread modeling currently involves running hundreds of millions of wildfire spread simulations to understand the distribution of scenarios, including the worst case~\cite{worstcase}.  This type of modeling is outside the paper's scope and we leave it to other modeling efforts.} 

We find that, after {the distribution feeder (referred to simply as a ``feeder'' later in the paper)} length, weather and vegetation-related factors are most important for prediction in our study area (the service territory of PG\&E). Using accurate weather forecasts could further improve risk model performance, particularly for the transmission models we study. Derived weather { and vegetation} features (e.g. burn index, energy release content and vegetation height) improve distribution ignition risk models significantly, pointing toward the importance of improvements to derived weather and vegetation-related data models. Infrastructure data, including support structure and transformer age and conductor material, make small but meaningful improvements to model performance, and we encourage more utilities to make these infrastructure data public to enable researchers to develop improved predictive models.

\section{Data Description and Pre-processing}\label{Sec:data}
All data were recorded within PG\&E's territory between 1/1/2015 and 12/31/2019.  Table~\ref{tab:featuresummary} summarizes the resolution and types of data, and Table~\ref{abb} defines each abbreviation.  Figures 1 and 2 show the maps of non-confidential feeders and transmission lines in PG\&E. {Feeders are lines that transfer electricity from substations to consumers; their voltage levels are usually below 35 kV.  Transmission lines transfer electricity from the generation stations to substations; their voltage levels are typically larger than 60 kV.}

\begin{table*}[t]
\centering
\caption{Feature Summary}
\begin{tabular}{lm{11cm}m{3cm}}
\toprule
Feature Type   & Feature List & Resolution \\
\midrule
Vegetation     & Max Tree Height, Avg Tree Height   &  Yearly          \\
\midrule
Weather        & \parbox[t]{11cm}{\vspace{-.2cm}\textbf{gridMET:}  Max BI, Max ERC, Min ETR, Min FM100, Min FM1000, Min PET, Min PR, Min RH, Min SPH, Max SRAD, Max T, Max~VPD, Max WS \\ \textbf{Mesowest:} Max WS, Avg WS, Max GWS, Max T, Min RH, Distance from Weather Station to Feeder}          &  Daily          \\
\midrule
Infrastructure & \parbox[t]{11cm}{\vspace{-.2cm}\textbf{Feeder: }Voltage, Feeder Length, Tier~3\%, Tier~2\%, Zone~1\%, non-HFTD\%, Primary OH Conductor\%, Transformer Zone~R\%, Transformer Zone~S\%, Transformer Zone~T\%, Transformer Zone~X\%, Pole Material Wood\%, Pole Material THBR\%, Pole Material Steel\%, Conductor Wind Speed Code~2\%, Conductor Wind Speed Code~3\%, Conductor Wind Speed Code~4\%, Conductor Material AAC\%, Conductor Material ACSR\%, Conductor Material Copper\%, Avg Transformer Age, Avg OH Conductor Age, Avg Pole Age, Historical Ignition Count, Historical Wiredown Count, Distance to Weather Station\\ \textbf{Transmission Line:} Voltage, Line Length, Historical Ignition Count, Historical Wiredown Count, Avg Distance to Weather Stations}             &  \parbox[t]{3cm}{Constant except that all age-related features are yearly and historical wiredown and ignition counts are daily}   \\
\bottomrule
\end{tabular}
\label{tab:featuresummary}
\end{table*}

\begin{table}[t]
\centering
\caption{Weather Feature Abbreviation}
\begin{tabular}{llll}
\toprule
BI    & Burning Index       & ERC    & Energy Release Component           \\
ETR & Evapotranspiration    & PET & Potential Evapotranspiration  \\
FM100 & Fuel Moisture 100hr & FM1000 & Fuel Moisture 1000hr      \\
PR    & Precipitation       &   SRAD  & Shortwave Radiation           \\
RH  & Relative Humidity        & SPH & Specific Humidity  \\
 T      & Temperature              & VPD & Vapor Pressure Deficit       \\
WS    & Wind Speed          &   GWS     &    Gust Wind Speed                              \\
\bottomrule
\end{tabular}
\label{abb}
\end{table}
\subsection{Infrastructure Records}

\begin{figure}[t]
	\centering	        	
    \includegraphics[width=.65\linewidth]{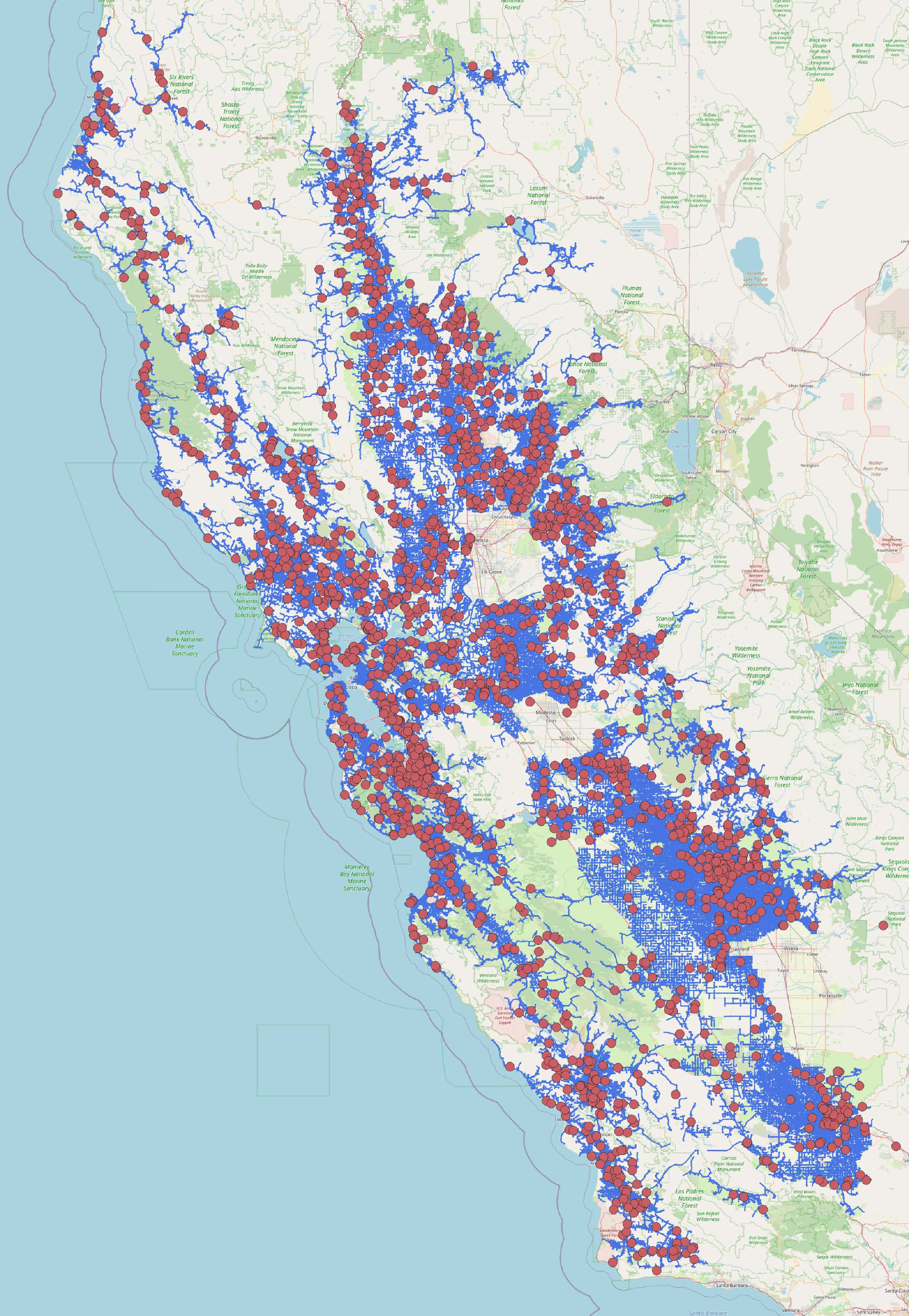}
    \caption{The location of distribution feeders in PG\&E, including ignitions (red circles) happened from 2015-2019.}
    \label{fig:feeder}
\end{figure}

\begin{figure}[t]
	\centering	        	
    \includegraphics[width=.65\linewidth]{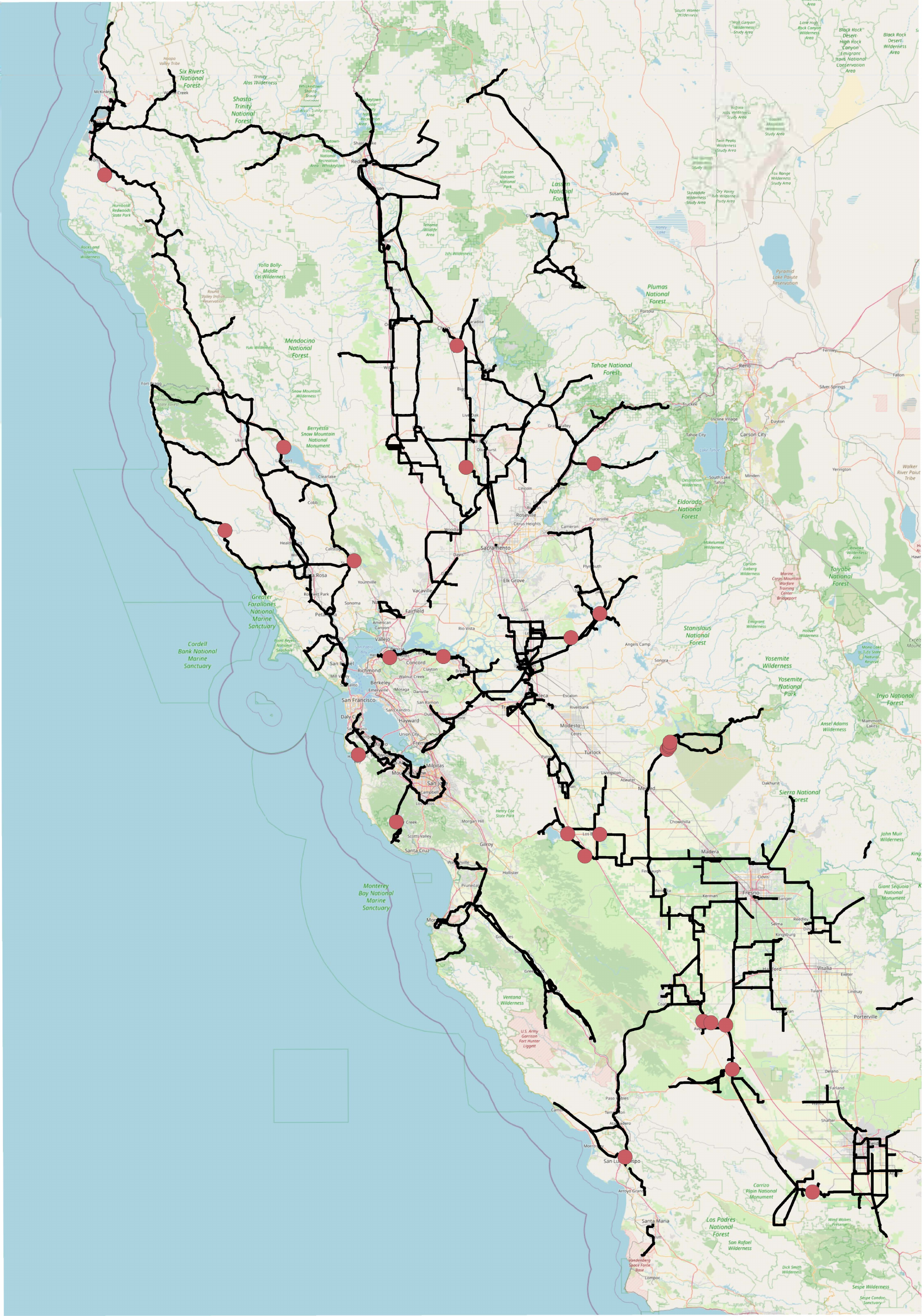}
    \caption{The locations {of 60-70 kV} transmission lines in PG\&E, including ignitions (red circles) from 2015-2019.}
    \label{fig:TL}
\end{figure}

\subsubsection{Distribution Feeders}

We merged the infrastructure information from different data sources by distribution circuit ID. These data include voltage level, and infrastructure component characteristics obtained from PG\&E's \textbf{EDGIS2-12.gdb} database~\cite{EDGIS_source}; feeder territory location information from PG\&E's \textbf{SDR Spreadsheet}~\cite{SDR_xlsx}; and conductor characteristics from PG\&E's \textbf{MGRA.gdb} database~\cite{datarequest}.  
Our data also include the lengths (in miles) of each circuit in High Fire-Threat Districts (HFTD) zones (Tier 3, Tier 2, Zone 1, or non-HFTD), taken from the \textbf{SDR Spreadsheet}. { These areas are referred to as Tier 3, Tier 2, Zone 1, and non-HFTD Zones, with Tier 3 being highest risk and non-HFTD being the lowest risk.}

For each transformer, support structure (pole), and overhead conductor in the catalog, the raw data tables in \textbf{EDGIS2-12.gdb} and \textbf{MGRA.gdb} provide ‘Installation Date’ and/or ‘Installation Job Year’ fields.  We aggregate these data to the feeder level by averaging\footnote{{ We found that including maximum age features does not improve the model so we opted to leave maximum age out and to focus on average age instead.}} the age of all similar components.  The raw data also include fields describing support structure material, transformer climate zone
conductor material and wind speed code. { In its raw form, wind speed code is a feature that takes on integer values between 1 and 4, inclusive.  While PG\&E did not provide documentation on the meaning of this feature, it appears to correspond to the 4-category Wind Speed Zones from ASTM E1996~\cite{speedcode}, with Category 1 conveying relatively low maximum wind speeds, and Category 4 corresponding to very high maximum wind speeds.} Climate zone is represented as the percentage of a given feeder's transformers in each of the four climate zones. Materials are coded as the fraction of each feeder's conductors or support structures that belongs to a given material class, and wind speed code is represented as the percentage of each conductor within each class. 

\subsubsection{Transmission Lines}
We obtained all geospatial information for transmission lines from the \textbf{ETGIS\_2-12.gdb} dataset (available at \cite{EDGIS_source}).  These files provide only location, voltage level, and line length for each transmission line below 70 kV in PG\&E's service~territory.

\subsection{Ignition and Wire-down Records}
Historical utility-induced ignitions are available in PG\&E's \textbf{IGNITIONS.gdb} database~\cite{SDR_xlsx}, { and the red dots in Figures~\ref{fig:feeder} and \ref{fig:TL} are the locations of these ignitions from 2015 - 2019. We assigned ignitions to a feeder or transmission line if the coordinates of the ignition fell within a 100-m buffer around the line. In rare cases when an ignition falls in more than one buffer, we assign the ignition to the closest line.}  We classify events according to the \textbf{IGNITIONS.gdb}'s listed cause: (1) Contact from vegetation; (2) contact from 3rd party (e.g. human, animal, car); (3) Equipment failure; (4) Unknown.  We also obtained historical wire-down events from the \textbf{SDR Spreadsheet}; each record includes the date and feeder or transmission line for the event.

\subsection{Aggregating Weather and Vegetation Data on Feeders and Transmission Lines}

We collected weather {feature}s from a simulated gridded meteorological dataset with daily resolution data on a 4-km grid (gridMET)~\cite{abatzoglou2013development} and the Mesowest dataset, which includes hourly meteorological data from specific weather stations~\cite{horel2002mesowest}. We also obtained tree height data at 10-m resolution from the Forest Observatory~\cite{ForestOb} data, which are synthesized from  remote sensing data. { (Note that because it is not directly measured, we refer to tree height as a derived feature.)} 

We assigned all weather points falling within a 3-km buffer to the associated feeder or line. We then took the maximum, minimum, and/or mean values of each weather field within the buffer, see Table~\ref{tab:featuresummary}.\footnote{ We selected the maximum or minimum of each gridMET feature by evaluating which extreme correspond to greater risk. For example, high wind speed and low humidity are important for fire ignition, so we used only max wind speed values and minimum humidity values. }
{Whereas the gridMET data have daily resolution, the temporal resolution of the Mesowest data is hourly. In order to have the same temporal resolution as the gridMET data, f}or each Mesowest weather station, we calculated the daily average and maximum values of wind speed, gust wind speed, and daily maximum value of the temperature and daily minimum value of relative humidity, and we paired each feeder with the closest Mesowest weather station. We created a 20-km buffer along each transmission line and averaged all { aforementioned maximum, minimum or average Mesowest weather features} within the buffer. For feeders we include the distance from the centroid to the closest weather station; for transmission lines we include the average shortest distance to weather stations. 

We used the Forest Observatory data to identify all tree data points within a 10-m buffer (to match the Forest Observatory grid) of a feeder or transmission line, and then calculated the average and maximum tree heights for each buffer. Tree height data are generated yearly, beginning in 2016; we associated 2016-2019 ignitions with the corresponding Forest Observatory year, and 2015 ignitions with 2016 Forest Observatory data. 

We merged the weather and vegetation data with the infrastructure data using feeder circuit IDs or transmission line names. Each record of the resulting combined dataset represents all features of one feeder or transmission line on a calendar day. We also recorded the total number of historical ignitions and wire-down events that happened on that circuit prior to that day.  We added binary fields indicating whether or not an ignition or wire-down happened on the day in question.  We relate ignition and wire-down events to the weather on the day when the event happened.  We explore the importance of weather forecast accuracy in Section~\ref{Sec:result}. 

We drop records when any of their fields were missing data, including 25 records (0.02\%) from the transmission data set and 1,764,437 records (41\%) from the feeder data set. Such a large amount of data was omitted from the feeder data primarily due to weather stations with poor wind gust data records.  Though 99.9\% of dropped feeder records are on days without ignitions, we removed 608 records on days with ignitions (24\% of all ignitions). 
The data set has 114,392 samples from 69 transmission lines, and 2,533,967 samples from 2,097 feeders.

\section{Methodology}\label{Sec:method}

We { provide a list below of the two basic types of binary classification algorithms in this paper. Note that we assume the ignition event as the positive class for distribution system models.  However, as shown in Figure~2, only 26 ignitions happened on 60-70 kV transmission lines in 2015 - 2019, and as a result, it is impossible to develop a risk model that performs well for transmission ignitions using current data. However, wire-down events are critical precursors to ignitions, and therefore we use wire-down events as the positive class for transmission lines.}

\begin{itemize}
    \item \textbf{Logistic Regression:} Logistic Regression (LR) is the generalization of linear regression~\cite{wright1995logistic}. LR uses a logistic function to predict the probability of the positive class. The coefficients of the model are estimated by maximizing a likelihood function. To avoid over-fitting, we employ L1 regularization.  We also normalize the input features so they will have zero mean values and standard deviations of 1.
    \item \textbf{Classification Tree:} We use Random Forests (RF)~\cite{breiman2001random} and Histogram-based Gradient Boosting classifiers (HGB)~\cite{ke2017lightgbm}. The Gini index impurity-based criterion is used to grow each tree. With these methods, the probability of the positive class is, for a given choice of features, the number of trees of the positive class divided by the total number of trees in the ensemble.

\end{itemize}

The imbalance degree (defined as $N^{+}/N^{-}$, where $N^{+}, N^{-}$ are the number of positive and negative records, respectively) is very low in our data: {0.16\%} for the transmission line wire down data and {0.06\%} for the feeder ignition data. We explored the following methods to improve the imbalance distribution of samples. 

\begin{itemize}
    \item \textbf{Data resampling}: 
    In this context, resampling has the aim of making the number of data points in the positive class equal to that of the negative class. Oversampling randomly duplicates samples in the positive class; undersampling randomly deletes samples in the negative class; and the Synthetic Minority Over-sampling Technique (SMOTE)~\cite{chawla2002smote}, interpolates the samples of the positive class to generate new samples of the positive class.
    \item \textbf{Class weight modification}: This method increases the weight of the positive class in the cost function, which more heavily penalizes the misclassification of positive class records to improve the positive class accuracy.
\end{itemize}

We used the area under the Receiver Operator Characteristic (ROC) curve~\cite{fawcett2006introduction} (AUC)
to evaluate classifier performance in cross-validation and testing, since it is not biased to the majority or minority class~\cite{he2013imbalanced}. The horizontal axis of the ROC is the false-positive-rate and the vertical axis is the true-positive-rate; each point on the curve corresponds to a different probability threshold used by the classifier to assign an event as positive or negative.

\section{Model Training}\label{Sec:result}
In this section, we train the risk model with different classification algorithms and imbalance strategies and evaluate the performance of each model. We split the samples into train and testing data by year. We envision this type of model being used to calculate future risk, and therefore the training data includes samples in 2015 - 2018 and the testing data includes samples from June to November in 2019{; we chose these dates because they are in the future relative to the training data, and because they are during Northern California's wildfire season, when it is most important to accurately predict ignition.} We developed all models using classifiers from the \texttt{sklearn} toolbox~\cite{scikit-learn}. For each model, we used 3-times 10-fold stratified cross-validation to tune the hyperparameters. 
In LR model, the solver `liblinear' is used, and the hyperparameter `the inverse of regularization strength' is tuned by the cross-validation. Table \ref{tab:hyperparam} summarizes the hyperparameters tuned by the cross-validation for the RF and HGB models. Please see our Github repository for more details.

\begin{table}[t]
\centering
\caption{The Hyperparameters of RF and HGB models}
\begin{tabular}{m{4cm}m{3cm}}
 \toprule
RF & HGB \\
\midrule
\parbox[t]{4cm}{\vspace{-.2cm}  The minimum number of samples required to split an internal node; The minimum number of samples required to be at a leaf node;  The number of features to consider when looking for the best split}          &  \parbox[t]{3cm}{The learning rate of boosting; The maximum tree depth; The maximum number of tree leafs; The maximum number of bins used in histogram}         \\
\bottomrule
\end{tabular}
\label{tab:hyperparam}
\end{table}

\subsection{Feeder Risk Model}

\begin{figure}[t]
\centering
\subfloat[]
  {\includegraphics[width=0.48\linewidth]{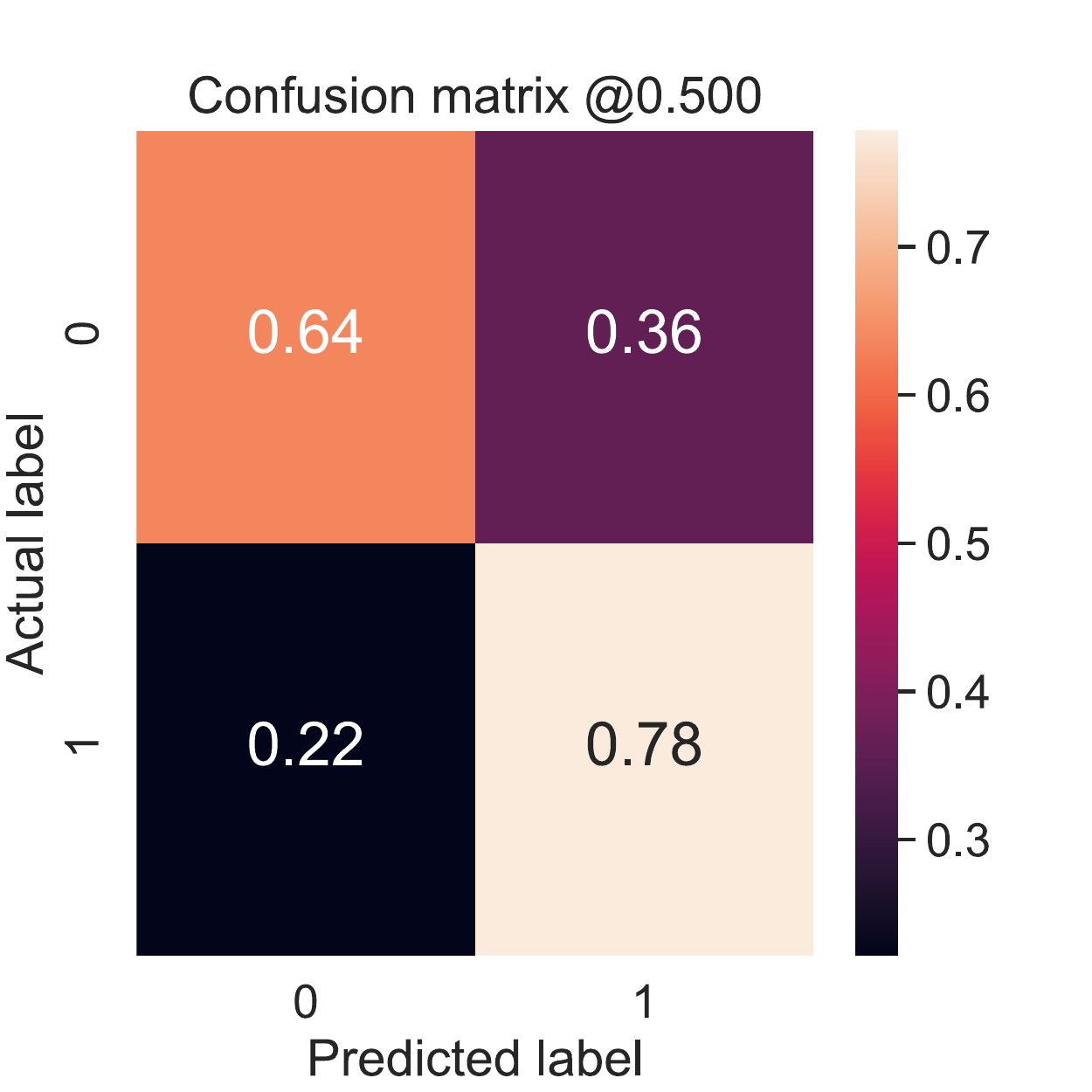}\label{fig:feeder_cm}}
  \hfill
\subfloat[]
  {\includegraphics[width=0.46\linewidth]{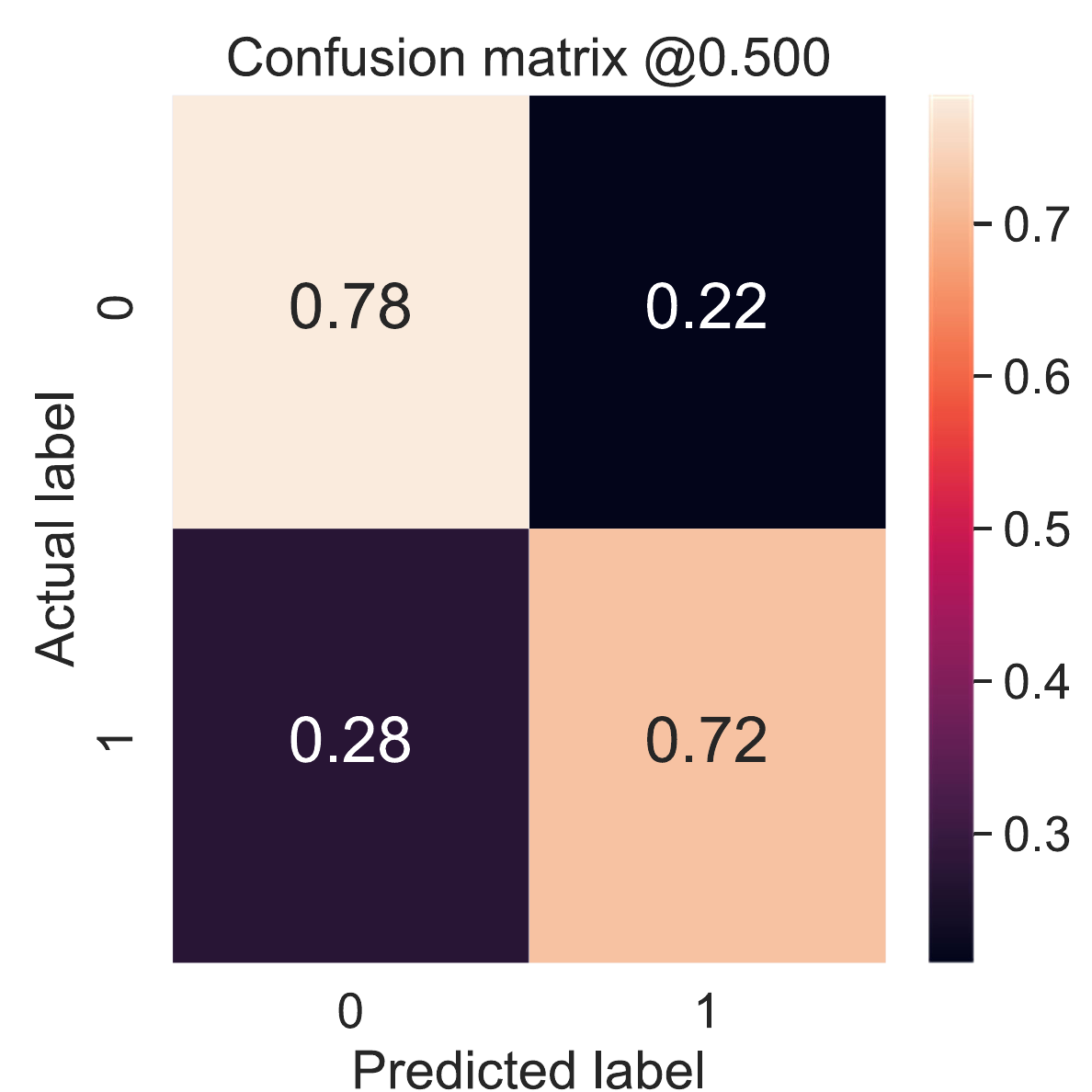}\label{fig:TL_cm}}
\caption{Confusion matrix when the threshold is 0.5. (a) Feeder ignition model. (b) Transmission wire down model. }
\vspace{-1em}
\end{figure}

Table \ref{tab:AUC_1} summarizes the mean and standard deviation (std) values of AUCs for cross-validation data and the AUC for the testing data. We considered the following two imbalance strategies: data resampling and class weight modification. Over-sampling, under-sampling, and SMOTE are realized by using the packages in \texttt{imblearn}.
The results in Table \ref{tab:AUC_1} are for the best models we obtained after cross-validation and hyperparameter tuning. {Note that the test AUCs are below the training AUCs; this is generally true for any machine learning model due to the presence of irreducible model error~\cite[p.34]{james2013introduction}. }
Overall, HGB with under-sampling shows the best test data performance, though balanced weight RF and LR also perform nearly as well on test data.
Figure~\subref{fig:feeder_cm} shows the confusion matrix when the threshold is 0.5.

\begin{table}[t]
\centering
\caption{AUCs for different feeder ignition models}
\begin{tabular}{ccccc}
\toprule
\multirow{2}{*}{Algorithm} & \multirow{2}{*}{Imbalance Strategy} & \multicolumn{2}{c}{Train} & \multirow{2}{*}{Test} \\
                           &                                     & Mean                 & Std                  &                       \\
                           \midrule
LR                         & -                                   & 0.670                & 0.028                & 0.668                 \\
LR                         & Under-sampling                       & 0.812                & 0.019                & 0.736                 \\
LR                         & Over-sampling                        & 0.833                & 0.019                & 0.756                 \\
LR                         & SMOTE                               & 0.831                & 0.019                & 0.756                 \\
LR                         & Balanced Weight                     & 0.833                & 0.019                & 0.757                 \\
\midrule
RF                         & -                                   & 0.823                & 0.017                & 0.757                 \\
RF                         & Under-sampling                       & 0.835                & 0.017                & 0.754                 \\
RF                         & SMOTE                               & 0.817                & 0.020                & 0.736                 \\
RF                         & Balanced Weight                     & 0.829                & 0.017                & 0.761                 \\
\midrule
HGB                        & -                                   & 0.822                & 0.021                & 0.753                 \\
HGB                        & Under-sampling                       & 0.837                & 0.018                & 0.776                 \\
HGB                        & Balanced Weight                     & 0.829                & 0.020                & 0.771                 \\
\bottomrule
\end{tabular}
\label{tab:AUC_1}
\end{table}

\subsection{Risk Model For Transmission Lines}
\begin{figure}[t]
	\centering	        	
    \includegraphics[width=0.4\textwidth]{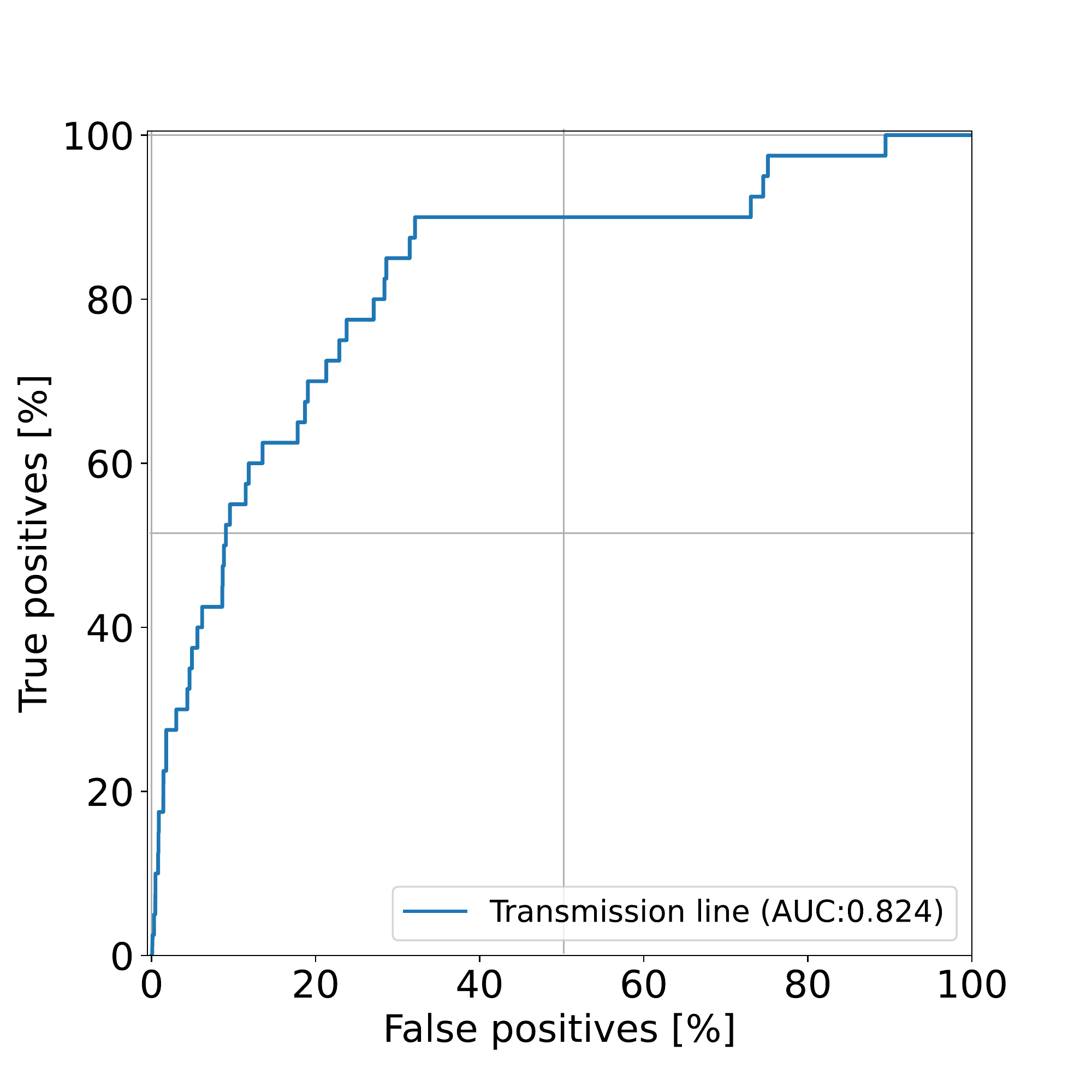}
    \vspace{-.25cm}
    \caption{ROC curve for the transmission line.}
    \vspace{-.5cm}
    \label{fig:TL_ROC}
\end{figure}

After following a process similar to that we used for the feeder model, we find the best model for transmission line wire down events is also the HGB with under-sampling. Figure~\ref{fig:TL_ROC} shows the ROC curve for the testing data; the AUC score is 0.824.  {The transmission model performs better than the feeder model; this is likely because wire down events are more frequent than ignition events.} Figure~\subref{fig:TL_cm} shows the confusion matrix when the threshold is 0.5. 

\section{Feature Importance Discussion}\label{Sec:dis}
Having maximized the performance of distribution and transmission ignition risk models, we now turn to investigating how different classes of data influence their performance.  Our ultimate objective is to understand what data are most valuable for ignition prediction and to provide recommendations for what types of data should be gathered, improved upon, and incorporated into risk models.  Though the data are specific to PG\&E's service territory, the process we employ can serve as a model for studies in other regions.

We begin by investigating the feeder risk models' feature importances. As we saw in the previous section, the HGB model's AUC is best, but the LR and RF models perform nearly as well.  Therefore we study and compare the performance of all models. For the HGB model, the importance is calculated by using the total gains of splits which use the feature~\cite{hastie2009elements}; the split gain is defined as the amount by which leaf node impurity improves with a given split. For the LR algorithm, we use the magnitude of the coefficients of the underlying linear model as measures of importance~\cite{ng2004feature}. For the RF algorithm, we use each feature's Gini importance~\cite{breiman2001random}.  These metrics do not indicate a \textit{causal} relationship, but rather measure each feature's correlation to the target variable and corresponding ability to improve model predictions.

\begin{figure}[t]
	\centering	        	
    \includegraphics[width=1\linewidth]{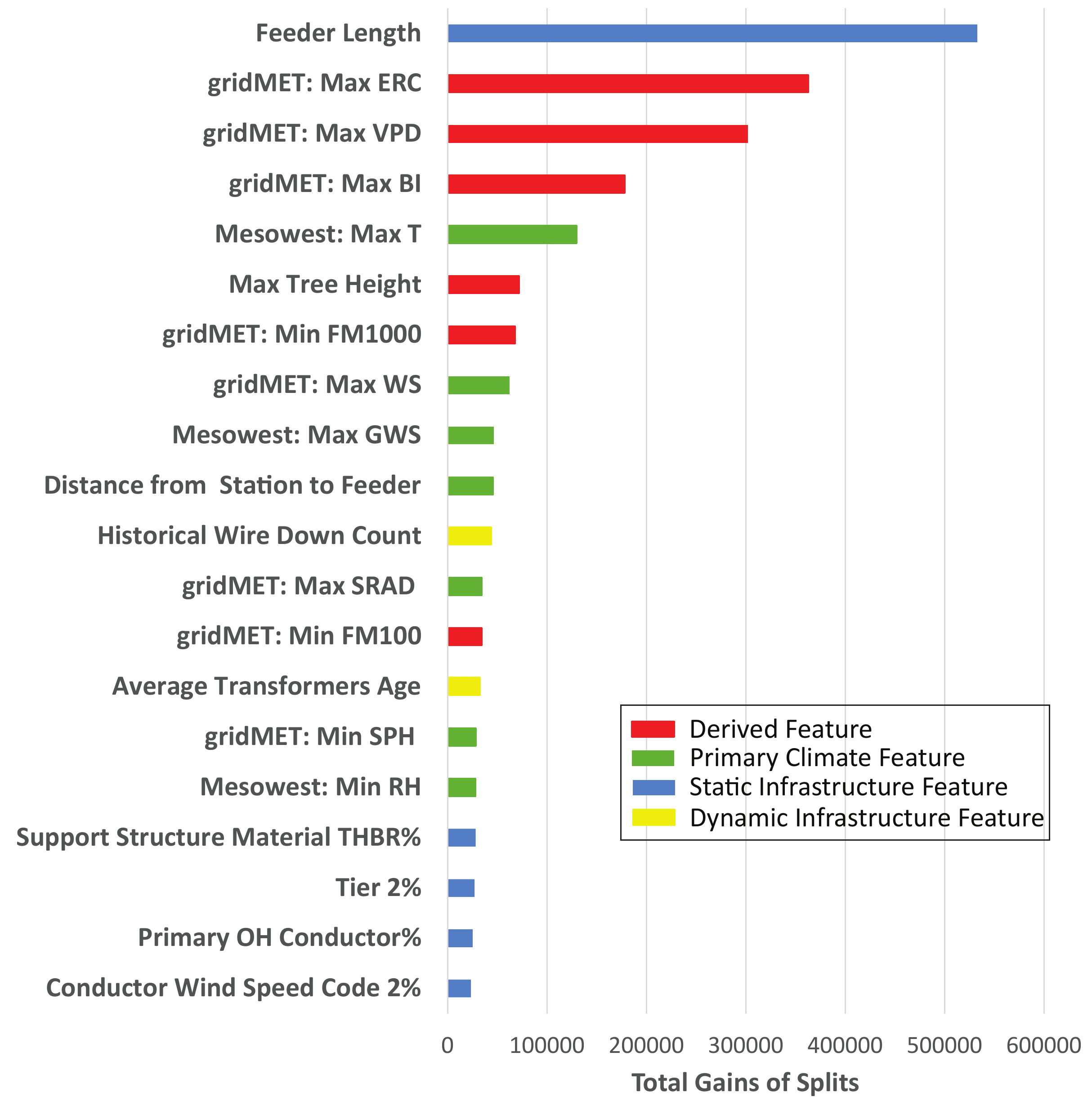}
    \vspace{-.25cm}
    \caption{Feature importance rank of the HGB model.}
    \label{fig:feeder_HGB_FI}
\end{figure}

\begin{figure*}[t]
	\centering	        	
    \includegraphics[width=\textwidth]{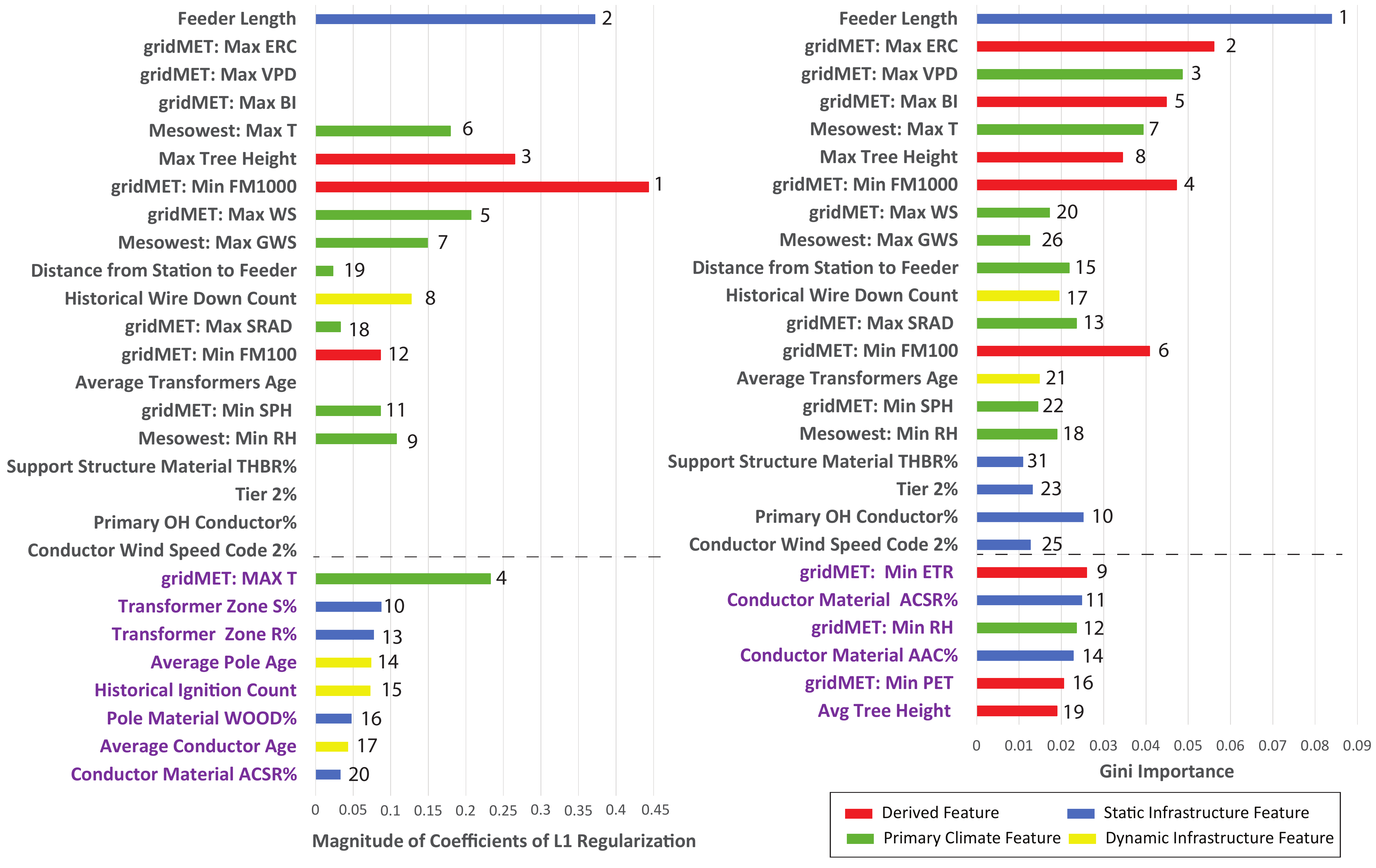}
    \vspace{-2em}
    \caption{Feature importance of the Logistic Regression model (left) and the Random Forest model(right). { The features in purple are those in the top-20 list for the LR and RF models but not in the top 20 of the HGB model.}}
    \label{fig:LR_RF_FI}
\end{figure*}

Figure~\ref{fig:feeder_HGB_FI} shows the feature importance rank of the HGB model, and Figure~\ref{fig:LR_RF_FI} shows the feature importance of the LR and RF models. To facilitate interpretation, we organize the features into four categories.  For climate and vegetation features we break measurements into \textit{primary climate features} and \textit{derived features}, as defined in the gridMET dataset~\cite{abatzoglou2013development}.  Primary climate features can be readily measured by weather stations (though in the case of gridMET these data are in fact generated via regional-scale reanalysis modeling) and are shown in green (PR, RH, SPH, SRAD, T, WS, GWS, and distance from weather station to feeder).  Derived features are not easily measured and are typically estimated with models and algorithms (in red; BI, ERC, ETR, PET, FM100, FM1000, VPD, and vegetation height).  For infrastructure features, we classify a portion of the data as \textit{dynamic infrastructure features}, i.e. those that require regular updating and may be specific to a given date (in yellow; transformer age, OH conductor age, pole age, and historical wire down and ignition counts).  We classify all remaining infrastructure features as \textit{static infrastructure features} (in blue).

In Figure~\ref{fig:feeder_HGB_FI} we investigate the feature importance for the best-performing HGB model with the 20 most important features. One can see that feeder length is the most important feature{\footnote{ To check the importance of the feeder length, we built a new model without `Feeder length', `Historical ignition count', and `Historical WD count'. (We also removed historical counts because they are highly correlated with feeder length). The AUC of this model reduces to 0.746. Apart from the obvious omission of these three features, the top-20 feature importance ranking is essentially unchanged in this new model.}}.  After feeder length, we find that derived features (red bars) {\footnote{The top derived features are ERC, VPD, and BI. ERC represents the potential heat release per unit area in the flaming zone; VPD is the deficit between the amount of moisture in the air and when air is saturated; BI is derived from the spread component and the energy release component. All these are highly related to wildfires.}} are among the most important predictors of ignition events. This suggests that efforts to maintain and improve model-generated climate and vegetation data are important in the continued development of wildfire risk models.  

Primary climate features provide limited additional predictive value to the HGB risk model.  This could be due to low data quality: Weather station data (mesowest) are often far from the feeder in question, and gridded data (gridMET) are available on relatively coarse spatial scales.  Features such as wind speed -- which exhibit strong spatial variance -- are particularly poorly represented by these datasets.  Accuracy would be even further reduced if one used forecast data; we will examine this issue later in this section.  These issues of data quality suggest that improving the spatial resolution and accuracy of certain primary climate data could potentially yield meaningful improvements to risk model performance.  On the other hand, primary data may be { less common} in the HGB feature importance list simply because other features more strongly correlate with ignition risk, regardless of the relative quality of the data.  Examining the top features in Figure~{\ref{fig:feeder_HGB_FI}} one can see the most important {ones} capture the propensity for vegetation to burn once exposed to a spark.  Though the models we employ do not identify causal relationships, it may be that the most important underlying causal factors for ignition relate to propensity for fuel to burn, as opposed to the propensity for a spark to be generated in the first place.

Features describing infrastructure condition do not improve model performance to the same degree that weather features do, with the exception of feeder length.  This is surprising because one might reasonably expect asset condition to be associated with its propensity to fail and generate a spark. We find that factors that can be associated with drivers of infrastructure damage -- including transformer age, pole material, and conductor wind speed code -- are eclipsed in importance by other primary climate and derived features.  However, we note that we do not have access to electric compliance tags However, we note that we do not have access to electric compliance tags ({e.g., maintenance date, inspection date, replacement record,} which indicate a potential issue with the condition of a piece of infrastructure) in this study; these may provide stronger predictive value than other infrastructure data and we encourage utilities to release this class of data for researchers to improve prediction models.  We also note that historical wire-down counts show valuable importance, but in no case is this dynamic infrastructure feature as valuable as the primary climate and derived features.  This indicates the tremendous complexity of electricity ignition risk prediction; it is not simply that there are ``problem circuits'' that repeatedly fail.  

In Table~\ref{tab:feature} we show model performance when we limit the features available for training to specific categories.  These results corroborate our prior observations based feature importances: derived climate and vegetation data on their own provide significantly more predictive capacity than primary climate data can in isolation, and combining the two classes of data only marginally improves model performance over the derived-data-only model.  

The observation that weather- and fuel-related features dominate the feature importance lists is consistent with PG\&E's strategy of identifying circuits for {Public Safety Power Shutoff (PSPS)} by first evaluating which regions will experience weather and fire conditions beyond a threshold.  However we note that infrastructure and historical ignition data do provide \textit{some} predictive value, as indicated in both Figure~\ref{fig:feeder_HGB_FI} and Table~\ref{tab:feature}{, where the primary climate and derived model improves after adding infrastructure features.  Note that it is not suitable to use \textit{only} infrastructure data to train the model, since most infrastructure data are static and we are building models to capture daily variation in risk. Our results show that infrastructure features improve AUC in an amount comparable to derived features.} This suggests that integrating infrastructure data into initial screens for high risk feeders could improve prediction quality. We encourage more utilities to make  infrastructure characteristics publicly available for researchers to improve risk modeling; we also encourage continued auditing and improvement of data collection in the field.

\begin{table}[t]
\centering
\caption{AUCs when using different features for feeder}
\begin{tabular}{lccc}
\toprule 
\multirow{2}{*}{Feature} & \multicolumn{2}{c}{Train} & \multirow{2}{*}{Test} \cr 
                                                               & Mean                 & Std                  &                      \\ 
                           \midrule
Primary Climate  &  0.7902 & 0.019 & 0.698 \\
Derived   & 0.806 & 0.020 & 0.730\\
Primary Climate and Derived  & 0.822 & 0.018 & 0.740 \\
Primary Climate, Derived, and Infrastructure  & 0.839& 0.018& 0.776  \\
\toprule
\end{tabular}
\label{tab:feature}
\end{table}

\begin{figure*}[t]
	\centering	        	
    \includegraphics[width=0.9\textwidth]{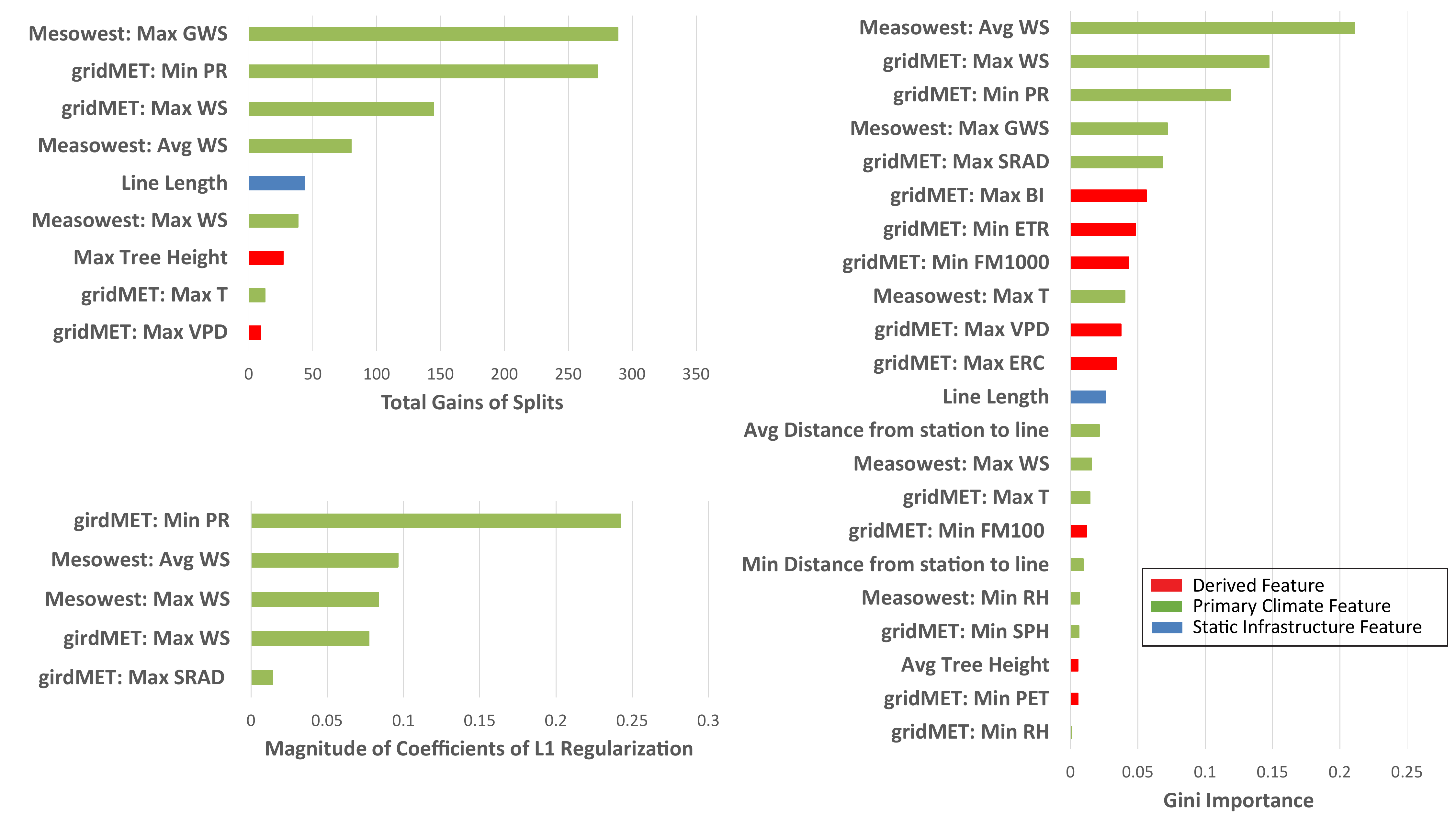}
    \vspace{-.3cm}
    \caption{Feature importance rank of the HGB (top left), LR (bottom left), and RF (right) models for the transmission line.}
    \vspace{-.5cm}
    \label{fig:FI_TL}
\end{figure*}

In Figure~\ref{fig:LR_RF_FI}, we explore feature importances for the LR and RF models. To have a better comparison with the importance rank of the HGB, we list features in the same order as in Figure~\ref{fig:feeder_HGB_FI}, and number the actual rank of each feature beside each bar. In addition to the top-20 important features of the HGB, features in the top-20 rank of the LR and RF models are shown in purple under the dashed lines. We observe that derived features that are important in HGB are removed as non-important features by the LR model's L1 regularization. Unlike the LR model, the RF model shares a more similar list of the top-20 important features with the HGB model -- possibly since they are both tree-based -- but the order is different. Looking at all three models' feature importance, we find that the three top-20 lists include all four categories of features, both static and dynamic infrastructure features are shown in the lists, but primary climate and derived features still dominate the top-10 important feature lists.  Importantly, { more primary climate features are listed in the top-20 ranks of the LR and the RF models, indicating that} primary climate features provide more predictive power in both the LR and RF models than in the HGB model.  This points to the importance of improving data quality in all classes where possible.  Derived features such as tree height and fuel moisture are consistently important across all models.  

To investigate the importance of weather forecast accuracy, we build an HGB model using the average values of weather features seven days prior to each event. Recall that our base case models above use \textit{actual} weather on the day of the event.  Comparing the two models in this way {(one with contemporaneous weather data, one with recent averaged weather data)} provides a range of performance that weather forecasts could provide, with the historical weather data model representing performance with a crude weather forecast and the actual weather data model representing performance if one had access to a {zero-error} weather forecast.  Using historical average weather data, the model AUC is 0.743, as compared to 0.776 for the {zero-error} weather model; as shown in Figure~\ref{fig:histweather}, the AUCs are very similar.  This echoes our findings in the feature importance analysis and in Table~\ref{tab:feature}.  Specifically, improvements in forecasts will most impact the quality of features with higher variance such as wind speed and temperature. However, these features are less important, and therefore improvements in their forecasts have relatively small impact on model performance.

\begin{figure}[t]
	\centering	        	
    \includegraphics[width=0.4\textwidth]{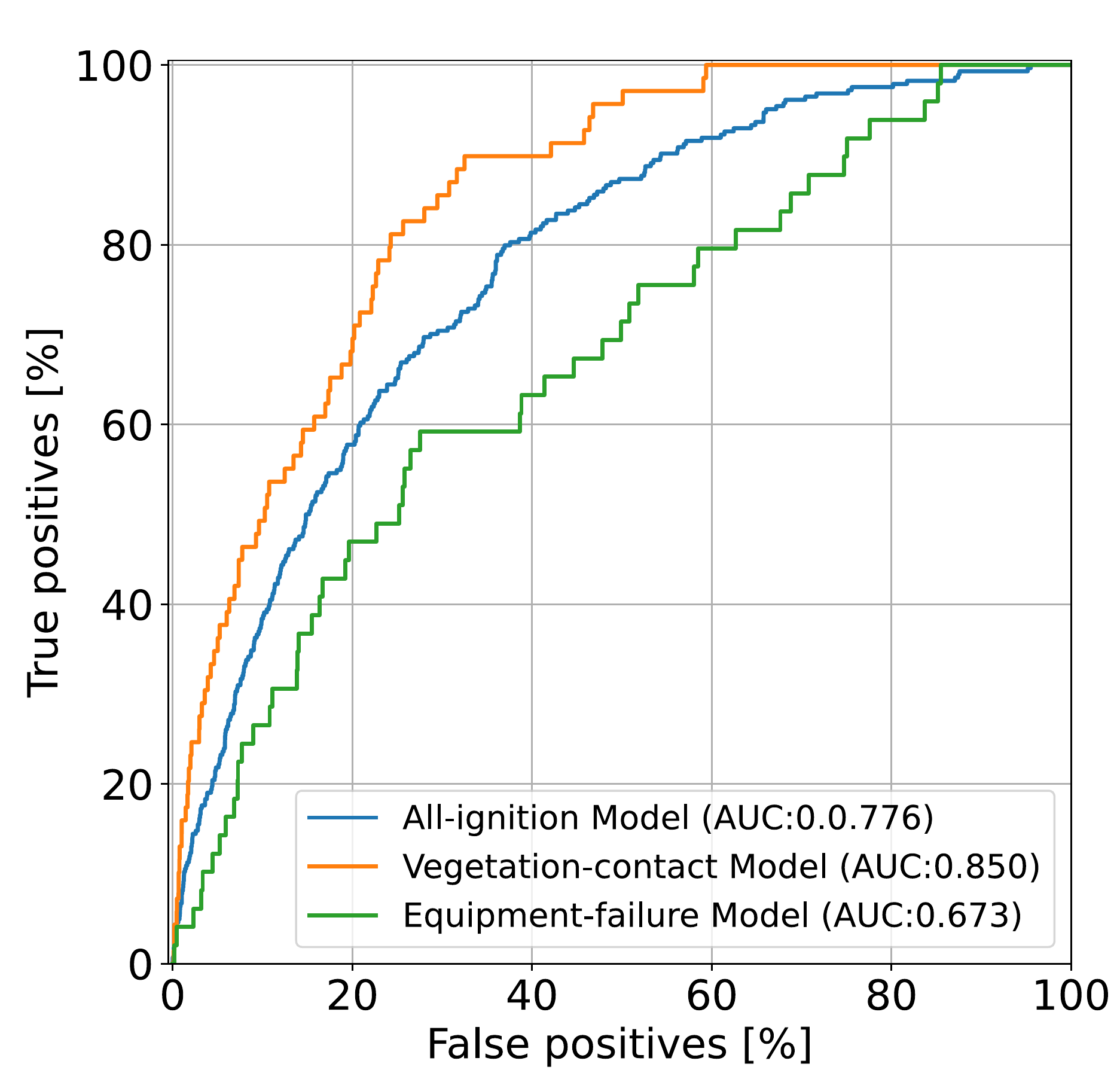}
    \vspace{-.3cm}
    \caption{ROC curves of the vegetation-contact model, equipment-failure model, and all-ignition model using the HGB-Undersampling model. The AUC scores are reported in the legend.}
    \vspace{-.5cm}
    \label{fig:VC_EF}
\end{figure}

\begin{figure}[t]
	\centering	        	
    \includegraphics[width=0.4\textwidth]{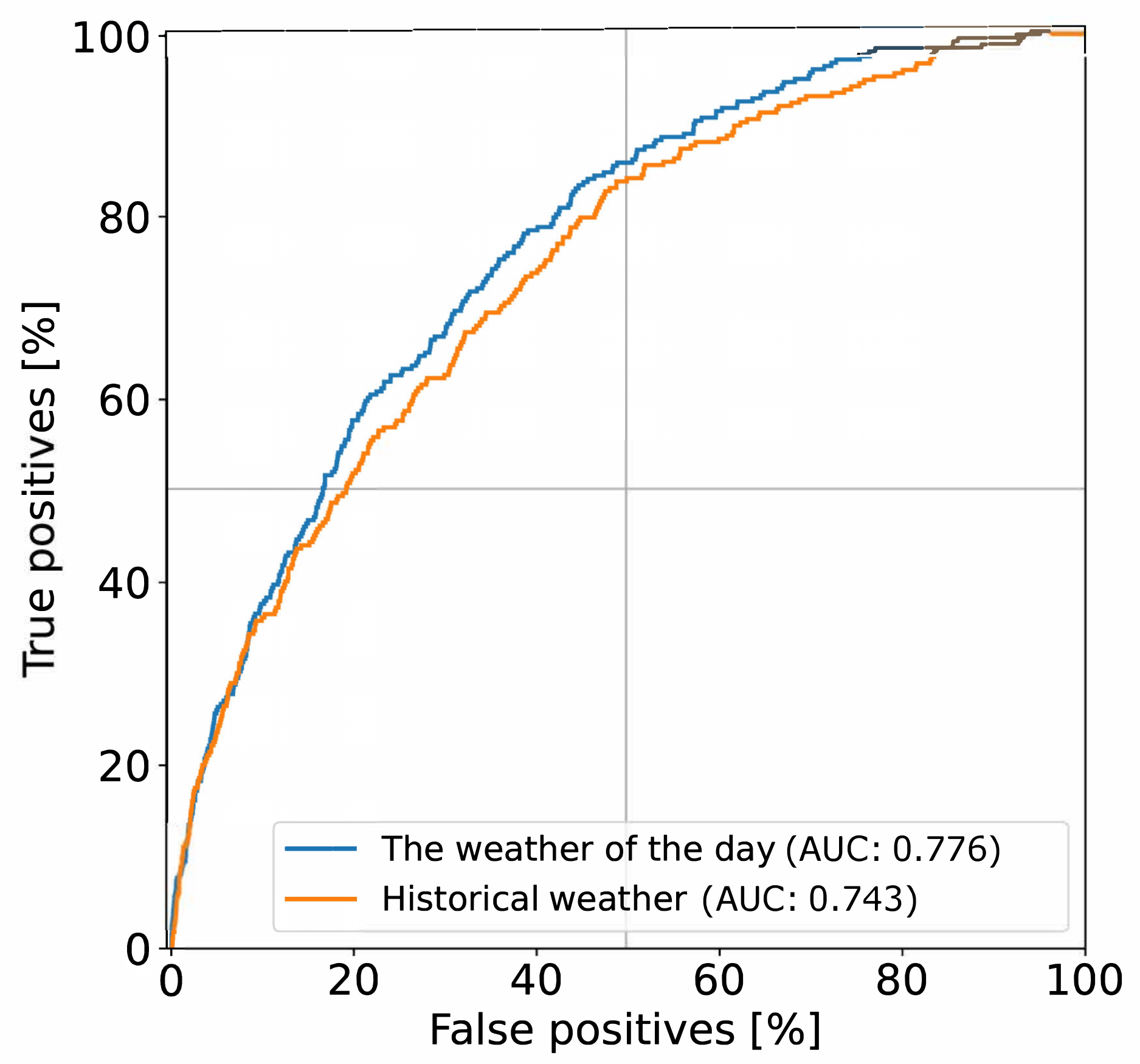}
    \vspace{-.3cm}
    \caption{ROC curves of the models using weather of the day and the historical weather before the day.}
    \vspace{-.5cm}
    \label{fig:histweather}
\end{figure}

We next {build two new} models that separately predict ignition risk due to vegetation contact versus equipment failure, and compare their performance to models that treat all classes of ignition in the same way (as above). When we train the vegetation contact ignition model, we label the days that had ignitions caused by other factors as negative, and we conduct similar procedure for the equipment failure ignition model. Figure~\ref{fig:VC_EF} shows the ROC curves for these two models as well as the all-ignition model using the HGB with under-sampling model. We observe that the vegetation-contact model is significantly better than the equipment-failure model.  We note that the vegetation-contact model achieves a true positive rate of 100\% at a much lower false positive rate than any of the other models.

Finally, we conduct a feature importance analysis for transmission line models. The feature importance ranks for each model are shown in Figure~\ref{fig:FI_TL} (the features with zero importance are not shown in the ranks).  Recall that, because ignition events are extremely rare for transmission lines at the voltage levels we studied, the transmission model uses wire down events as the target variable.   Unlike the feeder model, all models for the transmission line consistently show that \textit{primary} climate features -- especially wind speed and precipitation but also solar radiation -- are the most important features. This highlights the importance of high quality sensing and forecasting of these features for transmission line wire down prediction.  Moreover the dominance of wind speed in these models suggests the importance of wind-driven loading as a feature, possibly because it is a key failure mechanism.  Based on this evidence we speculate that electric compliance tag records are very important for generating high quality transmission line risk models.  

\section{Conclusion and Future Work}\label{Sec:conclusion}

By studying on historical PG\&E ignition and wire-down events, this work investigates what data are most valuable for risk prediction and thus provides recommendations for what types of data should be gathered, improved upon, and incorporated into risk models for other utilities and researchers. We find that weather and vegetation {feature}s are among the most important predictors of ignition events on distribution circuits and wire down events on transmission lines. Though both primary and derived climate data are important for ignition prediction on distribution circuits, derived climate variables (i.e. those generated with models from primary data, such as the burn index and vegetation energy release content) dominate our best performing model for distribution circuit ignitions.  In contrast, we find that primary weather variables are more valuable for transmission line wire down events, suggesting the importance of weather forecasting and meteorological station expansion near transmission lines.  

We found that extremely high resolution distribution infrastructure data provided small but measurable improvements to model performance.  Historical wire down data provided significant predictive value in some but not all models.  We encourage more utilities to release similar data sets for researchers to use to study risk model approaches and innovations.

The observation that infrastructure data do not provide the same level of predictive value as primary and derived weather and vegetation data conveys the depth of the challenge for utility operations.  Specifically, our results demonstrate that high risk conditions are not driven by key `problem circuits' or types of infrastructure.  Instead, in its present condition nearly all of PG\&E's infrastructure is at risk, and environmental variables that are outside utility control dominate measures of circuit risk.  This supports PG\&E's strategy of performing PSPS events based on weather and vegetation risk, and pursuing system-wide grid hardening.  

{ Taken together, our results lead us to conclude that that improvements to derived feature modeling will benefit distribution risk models; expanding meteorological station coverage is important for transmission risk models but less so for distribution; and that infrastructure data generate important improvements to risk models.  It is important to note that this study was possible only because infrastructure data were made publicly available by PG\&E.  Though such data are not commonly available for other utilities, there is increasing transparency among many utilities in the U.S. for infrastructure hosting capacity for new load and distributed generation~\cite{nagarajan2022data}.  These data use similar underlying data to those we used in this study.  We recommend concerted policy shifts among utilities -- particularly those in the fire-prone West -- to expand the scope of the data they release to include infrastructure data such as what PG\&E released.  }

Our ongoing work includes further reducing risk model false negative rates. {We plan to include more features in the model, e.g., terrain, normalized difference vegetation index (NDVI), downscaled wind data.} Deep learning algorithms may also further improve the models' performance. Moreover, risk model performance may depend on the spatial aggregation strategy; we expect that a grid-cell risk model (rather than a circuit risk model) would provide valuable insights on which part of feeders or transmission lines are most dangerous. Future work will explore the potential applications of our risk model; for example, the predicted ignition probabilities can be to inform the decision-making of wildfire mitigation problems.

\section*{Acknowledgements}
The data that support the findings of this study are openly available. This research was supported by the University of California Office of the President Laboratory Fees Program (Grant ID: LFR‐20‐652467).

\section*{Code availability statement}
The data that support the findings of this study are openly available. The code for the risk models is available on Github. The link to the Github repository can be found here: \url{https://github.com/MengqiYao/wildfireriskmodel}


%

\bibliographystyle{IEEEtran}
\bibliography{ref}

\end{document}